\documentclass[journal=ancac3]{achemso}   

\usepackage[utf8]{inputenc}
\usepackage[T1]{fontenc}
\PassOptionsToPackage{hyphens}{url}
\usepackage{hyperref,url}
\usepackage{amsmath}
\usepackage{amssymb}
\usepackage{amsthm}
\usepackage{physics}
\usepackage{subcaption}
\usepackage{float}
\usepackage{here}
\usepackage{ragged2e}
\usepackage[normalem]{ulem}
\usepackage{soul} 
\usepackage{graphicx}
\usepackage{dcolumn}
\usepackage[title]{appendix}
\usepackage{bm}
\usepackage{xcolor}
\usepackage[format=plain]{caption}
\usepackage{wrapfig}
\definecolor{mygreen}{rgb}{0.2,0.7,0.2}  
\graphicspath{{Figures/}}

\title{Building unconventional magnetic phases on graphene by H atom manipulation: From altermagnets to Lieb ferrimagnets}

\author{B. Viña-Bausá}
\affiliation{Departamento de F\'isica de la Materia Condensada, Universidad Aut\'{o}noma de Madrid, E-28049 Madrid, Spain}

\author{M. A. Garc\'ia-Bl\'azquez}
\affiliation{Departamento de F\'isica de la Materia Condensada, Universidad Aut\'{o}noma de Madrid, E-28049 Madrid, Spain}

\author{S. Chourasia}
\affiliation{Departamento de F\'isica de la Materia Condensada, Universidad Aut\'{o}noma de Madrid, E-28049 Madrid, Spain}

\author{R. Carrasco}
\affiliation{Departamento de F\'isica de la Materia Condensada, Universidad Aut\'{o}noma de Madrid, E-28049 Madrid, Spain}

\author{D. Exp\'osito}
\affiliation{Departamento de F\'isica de la Materia Condensada, Universidad Aut\'{o}noma de Madrid, E-28049 Madrid, Spain}

\author{I. Brihuega}
\email{ivan.brihuega@uam.es}
\affiliation{Departamento de F\'isica de la Materia Condensada, Universidad Aut\'{o}noma de Madrid, E-28049 Madrid, Spain}
\alsoaffiliation{Condensed Matter Physics Center (IFIMAC), Universidad Aut\'onoma de Madrid, E-28049 Madrid, Spain.}
\alsoaffiliation{Instituto Nicol\'as Cabrera (INC), Universidad Aut\'onoma de Madrid, E-28049 Madrid, Spain.} 

\author{J. J. Palacios}
\affiliation{Departamento de F\'isica de la Materia Condensada, Universidad Aut\'{o}noma de Madrid, E-28049 Madrid, Spain}
\alsoaffiliation{Condensed Matter Physics Center (IFIMAC), Universidad Aut\'onoma de Madrid, E-28049 Madrid, Spain.}
\alsoaffiliation{Instituto Nicol\'as Cabrera (INC), Universidad Aut\'onoma de Madrid, E-28049 Madrid, Spain.} 

\begin{document}
\thispagestyle{empty}
\maketitle
\newpage
\begin{abstract}
Engineering all fundamental magnetic phases within a single material platform would mark a significant milestone in materials science and spintronics, reducing complexity and costs in device fabrication by eliminating the need for integrating and interfacing different materials.
Here, we demonstrate that graphene can host all non-relativistic magnetic phases-namely, diamagnetism, paramagnetism, ferromagnetism, antiferromagnetism, ferrimagnetism, altermagnetism and fully compensated ferrimagnetism ---by using single hydrogen atoms as building blocks. Through precise manipulation of these atoms by scanning tunneling microscopy, we can experimentally create all such magnetic phases. Their different magnetic character is confirmed by density functional theory and mean-field Hubbard calculations. In particular, we show that the new magnetic paradigm known as altermagnetism can be realized, exhibiting directionally spin-split energy bands coexisting with zero net magnetization due to protecting spatial symmetries. It is furthermore possible to create fully compensated ferrimagnets, lacking these symmetries and therefore presenting unrestricted spin splitting of the bands, with a vanishing net magnetization which in this case is protected by Lieb's theorem. These findings put forward H-functionalized graphene as a versatile platform to design, build and study these new emergent magnetic phases at the atomic scale. 
\end{abstract}
\flushbottom

\section{Introduction}
A new class of magnetic materials, merging the advantages of both ferromagnetic and antiferromagnetic systems, holds the potential to revolutionize next-generation  spintronic devices. These type of materials, 
notably those known as altermagnets but also including a variant of ferrimagnets,
are characterized by zero net magnetization, similar to antiferromagnets, yet retaining the spin-split electronic band structures characteristic of ferromagnets. This unique combination minimizes issues related to stray magnetic fields, facilitating miniaturization and device integration, while allowing spin-polarized electronic transport without the need for external magnetic fields or relativistic effects. Moreover, these materials offer directionally dependent electronic properties, tunable magnetic features, and hold significant potential for faster operational speeds, lower power consumption, and enhanced thermal transport ~\cite{EmergingLandscapeAltermagnetism, Jungwirth2022, SpontaneousHall, Bai2024, zhou2024crystal}.
This unconventional magnetism also opens pathways to novel material phases with parity beyond s-wave symmetry, providing potential magnetic analogs to high-parity superconductivity~\cite{Schofield2009}.

The study of altermagnets, in particular, is rapidly progressing~\cite{EmergingLandscapeAltermagnetism, Jungwirth2022, SpontaneousHall, guo2023, egorov2021antiferromagnetism, Bai2024}. In 2024, driven by recent theoretical and computational efforts, different crystalline systems, including MnTe~\cite{NatureAltermagnetsA, Lee2024, Amin2024}, RuO$_2$~\cite{Fedchenko2024} and CrSb~\cite{Reimers2024} were reported as the first experimental realizations of altermagnetism. On the other hand, the combination of zero net magnetization and spin-split bands, by definition present in altermagnets, is shared by another class of magnetic systems---the fully compensated ferrimagnets. The difference between both is that the zero net magnetization of the former is protected by space-time (or spin) symmetries, while in the latter it is instead achieved by fine-tuning the composition or the structure with external manipulations~\cite{finley2020spintronics,Cai2020, Siddiqui2018, Caretta2018, Kawamura2024, Nayak2015,xu2023evolution}. Until now, theoretical and experimental efforts have primarily focused on a materials-based approach---identifying or synthesizing compounds that inherently possess the desired properties,  hosted by a crystalline lattice with a suitable symmetry group upon magnetization. 

In this work, guided by spin groups symmetry selection, we propose an alternative
strategy to build unconventional compensated magnetic states—specifically altermagnets and compensated ferrimagnets---based on
an atom-by-atom manipulation.
Since spatial and spin symmetries determine the magnetic phases that can emerge, the capability to visualize and control 
them at the atomic scale is a key requirement of our bottom-up strategy. Scanning tunneling microscopy (STM) manipulation~\cite{Manipulation} provides such functionality, allowing for the strategic positioning of individual atoms on surfaces~\cite{Crommie1993, Hirjibehedin2006, Schneider2022}, thus enabling the construction of configurations with the appropriate symmetry. Atomically-resolved STM topographic images facilitate, in turn, the analysis of the existing spatial and spin symmetries.
We have selected hydrogen (H) atoms on graphene as a prototypical platform to test our approach~\cite{Gonzalez-Herrero2016}. This system combines several inherent properties that, put together, allow for the formation of all magnetic phases, conventional and unconventional, on a single material platform.

In brief, the proposed platform offers a) zero net magnetization, guaranteed by Lieb’s theorem, when there is an equal concentration of H atoms on both graphene sublattices,  b) manipulation capability of magnetic moments with atomic scale control, c) an anisotropic shape of the H-induced magnetic moment and, d) a long-range direct exchange interaction of both ferromagnetic and antiferromagnetic nature. With these ingredients at hand, we show in the following how all possible collinear, non-relativistic magnetic phases can be achieved by simply adjusting the 
arrangement of up to 4 H atoms in graphene.
Additionally, a highly symmetric altermagnetic phase can be realized with some particular arrangements involving 6 H atoms.

\section{Results}
\subsection{Designing magnetic phases atom by atom via symmetry selection}
Our approach consists in using the magnetic moments induced by single H atoms on graphene as building blocks to construct, following spin group symmetry considerations, all magnetic phases within a single material. In the following, we describe the magnetism induced in graphene by H atoms and introduce the essential symmetry rules driving the construction of magnetic phases (a thorough examination of this last topic can be found in the Discussion section).

The adsorption of a single H atom on graphene induces a magnetic moment without the need of external, intrinsically-magnetic, elements~\cite{Yazyev2010,Han2014}. The covalent bond between the hydrogenic $s$ orbital and the carbon $p_z$ (out-of-plane) orbital effectively removes an electron from the corresponding graphene sublattice, leaving an unpaired electron and therefore generating a  magnetic moment mostly localized in the complementary graphene sublattice. The spin cloud extends over several nanometers, exhibiting an anisotropic, triangular shape.  The induced magnetic moments couple ferromagnetically (antiferromagnetically) for H atoms adsorbed on the same (complementary) sublattice, and are essentially collinear, aided by the low strength of SOC in graphene. More generally, as follows from (the second) Lieb’s theorem for half-filled bipartite lattices~\cite{Lieb1989}, in a system with $N_A$ and $N_B$ H atoms chemisorbed on each graphene sublattice, the total spin in the ground state must be $S=|N_A-N_B|/2$.  By individually manipulating H atoms with a STM tip, one can achieve a desired distribution of these atoms across both sublattices of graphene \cite{Gonzalez-Herrero2016}, which enables selectively tuning  the material’s magnetization.    Alternative ways of obtaining local magnetic moments in this system have also been established~\cite{lehtinen2004irradiation,YazyevPRB,JJPalacios2008,UgedaPRL,Catarina2023}, in particular involving defects or suitable lattice terminations, which are essentially based on the same mechanism of generating unpaired electrons. However, engineering the magnetic moments with these approaches is, in practice, comparatively harder. 

Traditionally, magnetic systems have been described using (relativistic) magnetic groups, where symmetry operations act simultaneously on spin space and real space~\cite{mathematicaltheorysym, Tavger1956,yuan2021prediction}. However, in systems where the strength of spin-orbit coupling (SOC) is negligible (compared to the non-relativistic magnetization) additional symmetry operations appear, since those acting on the spin degrees of freedom are no longer coupled to the electronic coordinates. The corresponding framework is that of (non-relativistic) spin groups \cite{LITVIN1974538,Jungwirth2022, SpontaneousHall}, where symmetry transformations are applied independently on spin and real space,  enabling the prediction of features emanating from the magnetic ordering that are not accessible to magnetic groups \cite{EmergingLandscapeAltermagnetism}. In particular, they enforce the band structure to be centrosymmetric, that is $\varepsilon_{\sigma,\bm{k}}=\varepsilon_{\sigma,-\bm{k}}$ for $\sigma=\uparrow$ and $\downarrow$, irrespective of whether the magnetic system presents (spatial) inversion symmetry, hence predicting different spin textures in non-centrosymmetric magnetic materials. 

In simple terms, the symmetry-based procedure defined by this new theoretical framework to construct the different magnetic phases relies on identifying spatial transformations that return a given system to its initial configuration after performing a time-reversal operation ($\mathcal{T}$) on spin space, namely a spin-flip operation. For uncompensated magnetic systems, such as ferromagnets and general ferrimagnets, after carrying out a spin-flip operation there is no real-space transformation that can restore the original configuration. On the other hand, for compensated magnetic systems, there are different possibilities defining the existing magnetic phases. In antiferromagnets, spin degeneracy is enforced by an inversion or translation connecting opposite spin sublattices. Specifically, in two dimensional systems, the role of this spatial inversion can likewise be played by $C_{2,z}$: a $180^\circ$ rotation around the direction perpendicular to the plane of the lattice. In the novel altermagnetic phase, the system can be returned to its initial state by a real space rotation---or, as in our case, by a reflection through a vertical mirror plane---following  $\mathcal{T}$ acting on the spins~\cite{EmergingLandscapeAltermagnetism}. Finally, in fully compensated ferrimagnets, after carrying out a spin-flip operation there is no real-space symmetry restoring the initial configuration.

\subsection{Building conventional magnetic phases}
\begin{figure*}
\centering 
\includegraphics[width=\textwidth]{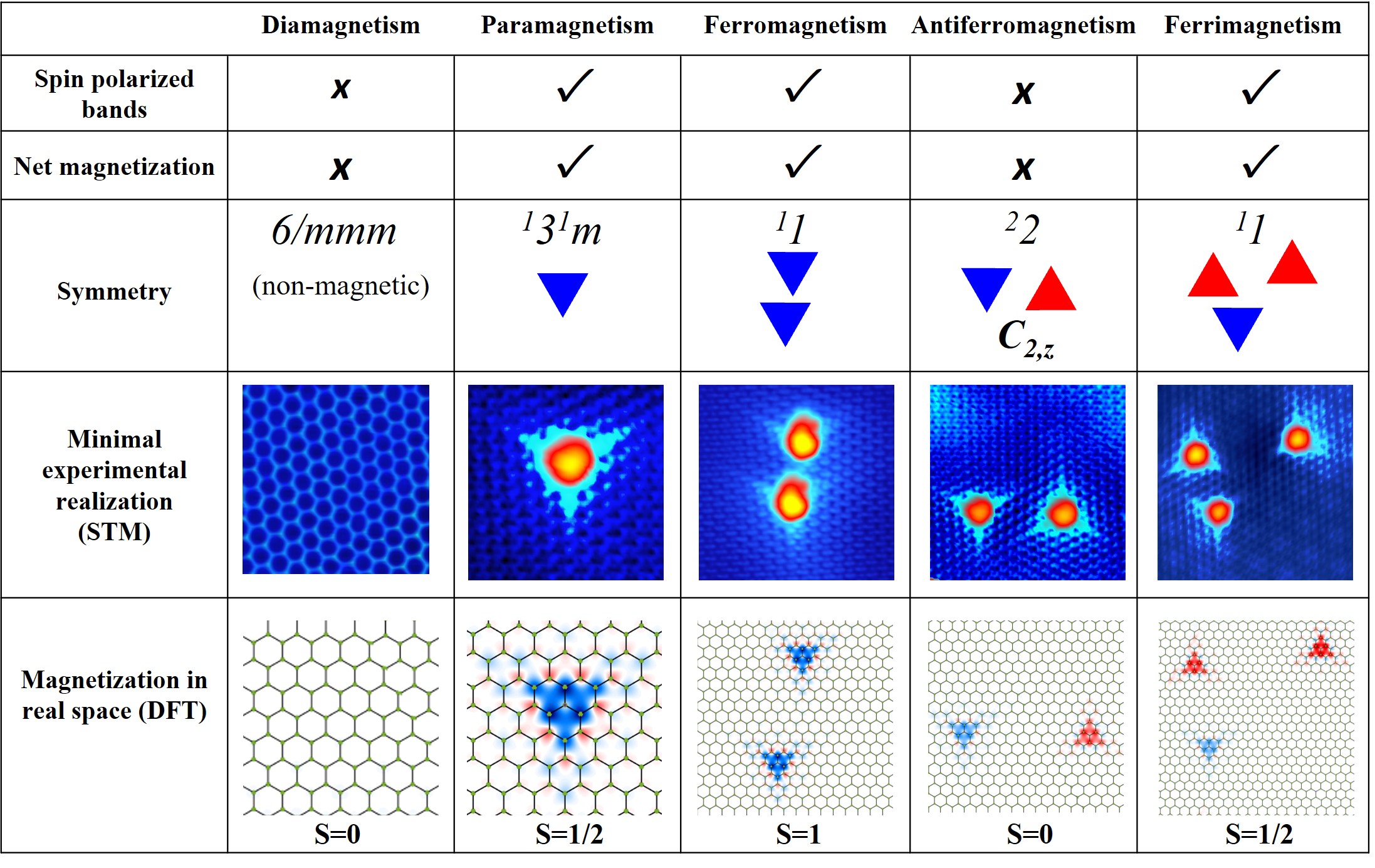}
\caption{\textbf{Building conventional magnetic phases using H atoms on graphene.} All possible non-relativistic, conventional magnetic phases can be realized on graphene by incorporating up to 3 H atoms. Point groups are indicated, in the spin group notation, for magnetic configurations presenting non-trivial symmetries. The two bottom rows show STM images and DFT-calculated magnetizations for each magnetic phase, clearly revealing that STM images encode the triangular magnetic orbital shape and the graphene sublattice adsorption site determining spin orientation. Representing each H atom as red or blue triangle (indicating opposite spin orientations) highlights the threefold anisotropy of the induced magnetic state. This symbolic approach enables the independent application of symmetry operations in both real and spin space, aiding in the design and identification of specific magnetic phases. STM image parameters by columns: (80mV, 0.1nA, 2.5x2.5nm$^2$); (50mV, 0.1nA, 4.7x4.7nm$^2$); (30mV, 0.1nA, 6.2x6.2nm$^2$ ); (40mV, 0.1nA, 7.5x7.5nm$^2$); (4mV, 0.05nA, 7.3x7.3nm$^2$).}
\label{Fig1}
\end{figure*}

As it was previously established~\cite{Gonzalez-Herrero2016, Yazyev2010}, and we here summarize in Figure \ref{Fig1}, all conventional magnetic phases can be formed with up to 3 H atoms on graphene: diamagnetism in the bare carbon lattice (0 H), paramagnetism with 1 H atom, ferromagnetism with 2 H atoms on the same sublattice , antiferromagnetism with 2 H atoms on complementary sublattices---as long as the separation between H atoms is larger than  $\gtrsim1.5$ nm \cite{palacios2014}---, and ferrimagnetism with 2 H atoms on one sublattice and 1 H atom on the complementary sublattice. A comparison with density functional theory (DFT) calculations reveals a clear correlation between the STM topography and the anisotropic, triangular-shaped distribution of the induced magnetic moment, see Figure \ref{Fig1} and Figure \ref{Fig2}a. For H atoms located on different graphene sublattices, the triangular magnetic distributions are predominantly of opposite spins and point in opposite directions, 
which enables the identification of the relative spin orientations in the experiment: parallel for triangles oriented in the same direction and antiparallel for those oriented oppositely. 

\subsection{Building unconventional magnetic phases}
In order to realize unconventional magnetic phases in hydrogenated graphene, it is necessary to consider a larger number of H atoms. In this work we mainly focus on 4 H arrangements, which represent the minimal configurations necessary to form both altermagnetic and compensated ferrimagnetic states. We show their magnetic properties and how we experimentally constructed the three magnetically compensated structures that can be formed in graphene: antiferromagnets, altermagnets, and Lieb (compensated) ferrimagnets. 

\begin{figure*}
\centering 
\includegraphics[width=\textwidth]{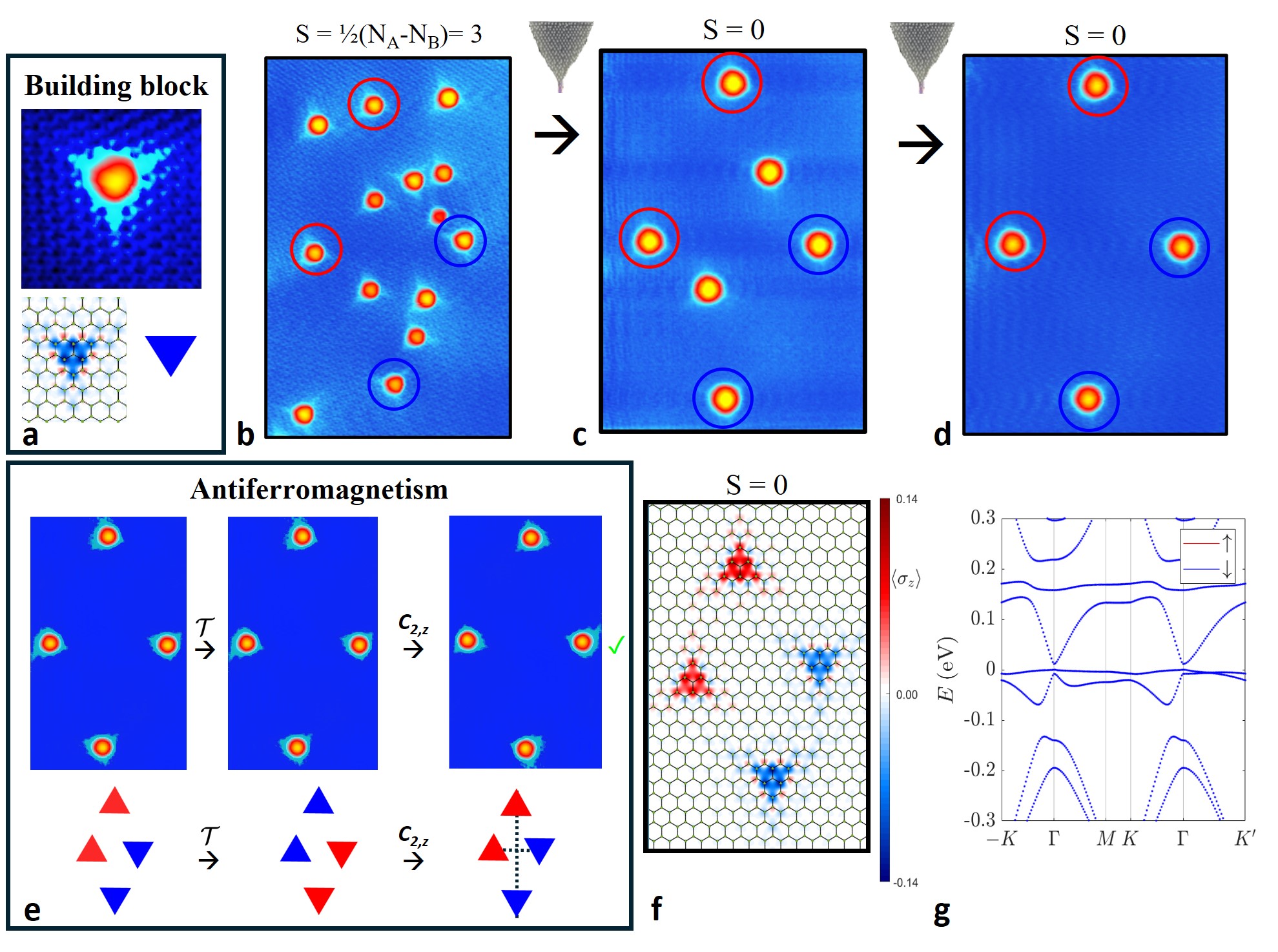}
\caption{\textbf{Building magnetic phases atom by atom. The antiferromagnetic case.} a) STM image (top) and DFT calculated magnetization (bottom left) of a single H atom on graphene. Schematized as a triangle (bottom right), it is proposed as a S=1/2 building block for engineering magnetic phases on graphene. b-d) Sequence of STM images of the same graphene region, where selected H atoms are subsequently removed to construct a 4 H parallelogram (to construct a $C_{2,z}-$symmetric structure corresponding to an antiferromagnet). e) Symmetry operations in spin and real space applied to both the experimental STM image and its symbolic representation. After the time-reversal operation, the twofold rotation brings the system back to its initial state. f-g) DFT-calculated magnetization and energy bands of a rescaled configuration showing a total spin S=0 and fully spin degenerated energy bands. STM image parameters: b) 40mV, 0.1nA, 15.2x24nm$^2$  c,d,e) 80mV, 0.1nA, 15.2x22nm$^2$.}
\label{Fig2}
\end{figure*}

As explained below in the Discussion section, with 4 H atoms the antiferromagnetic case must be realized with a parallelogram shape. In Figure \ref{Fig2}b-d, we show a sequence of STM images measured of the same graphene region, illustrating the manipulation of atomic hydrogen to construct such a 4 H parallelogram. Initially, multiple hydrogen atoms are gathered from the graphene surface onto the apex of the STM tip, see Methods for details. As shown in Figure \ref{Fig2}b, we subsequently deposit the hydrogen atoms onto a selected region of the graphene surface. While atomic-scale precision is not achieved during the deposition step, we control the final structure by selectively removing individual H atoms to achieve the desired configuration and symmetry. In this case, an antiferromagnetic parallelogram can be achieved by removing all H atoms except those marked with circles, where the red and blue colors denote the adsorption sublattices, corresponding to opposite spin orientations. Figure \ref{Fig2}c shows the same graphene region after we have selectively removed 8 H atoms to achieve a magnetically compensated configuration with equal number of hydrogen atoms on each sublattice ($N_A = N_B =3$). As guaranteed by Lieb's theorem, the region has a total vanishing net magnetization, while still lacking $[C_{2,z}\parallel\mathcal{T}]$ symmetry, where the operation to the left (right) of the double line $\parallel$ acts on real (spin) space exclusively (see Discussion section for more details). The final step is shown in Figure \ref{Fig2}d, where the remaining 2 H atoms have been removed to create the desired H parallelogram. 

The spatial symmetries are encoded in the local density of states measured in the topographic STM images, combining the contribution of both spin states. It is then easy to verify the spatial and spin symmetries from these measurements, recalling the well-established association: equal sublattice $\leftrightarrow$ equal magnetic moment orientation. The time-reversal (or spin-flip) operation $\mathcal{T}$ does not affect the appearance of the STM image, as it operates solely within the spin space (producing a spin-flip). The application of the $C_{2,z}$ rotation 
to the STM image in this case restores the system to its initial configuration, see Figure \ref{Fig2}e, and the product 
$[C_{2,z}\parallel\mathcal{T}]$ is therefore a symmetry of the system. Such symmetry enforces the global spin degeneracy of the bands (Kramer's degeneracy). This is confirmed by DFT calculations, see Fig. \ref{Fig2}f-g, displaying the characteristic zero net magnetic moment and fully degenerate energy bands. These calculations were conducted using a supercell approach to reveal band structures; in particular rescaling to a smaller size of $24\times24$ (1156 atoms) due to computational constraints, while keeping the proportions of the experimental geometry. Mean-field Hubbard tight-binding calculations are also presented in Supplementary Figure 1.

\begin{figure*}
\centering 
\includegraphics[width=380pt]{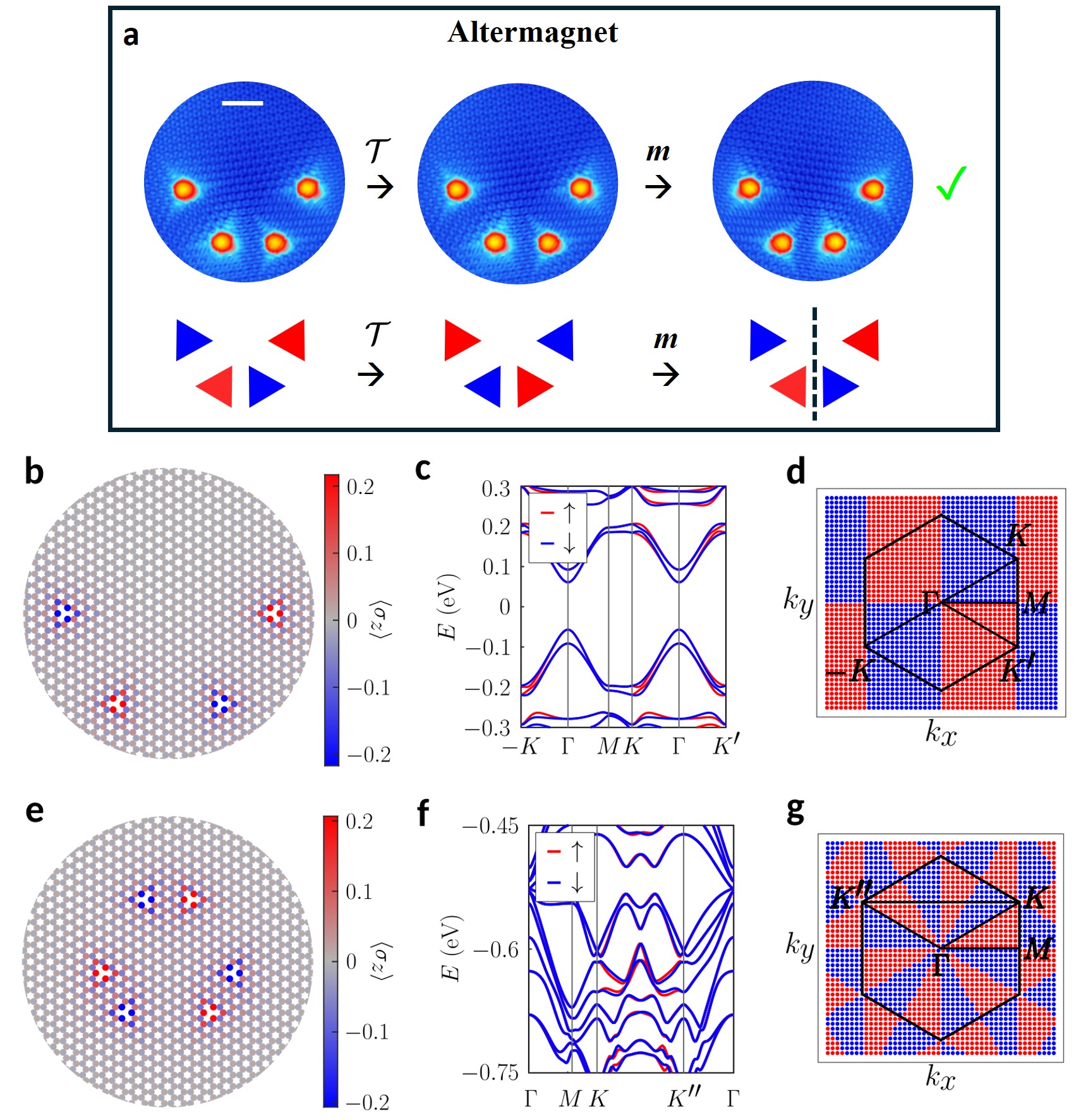}
\caption{\textbf{Hydrogenated graphene as an altermagnet.} a) Top: STM image (120mV, 0.1nA, scalebar=2nm) and symmetry transformations of the minimal experimental realization of an altermagnet, built, using the STM tip, by arranging 4 H on alternating graphene sublattices with a vertical mirror symmetry ($^2m$ spin group). Bottom: Triangle schematics and symmetry  transformations of the configuration. By applying $\mathcal{T}$ in spin space and a vertical mirror symmetry in real space we recover the original configuration b-c) Atomic magnetization and energy bands of a rescaled configuration corresponding to a), calculated using a mean-field Hubbard model. 
d) Spin texture of the highest occupied valence band in the b-c) configuration across reciprocal space, where the hexagon delimits the Brillouin zone and red (blue) color represents spin $\uparrow$ ($\downarrow$). The texture corresponds to a $d-$wave. e-f) Atomic magnetization and energy bands corresponding to a different configuration with spin group $^{1}3^{2}m$, calculated from a mean-field Hubbard model. g) Spin texture of a band in the e-f) configuration across reciprocal space, exhibiting an $i-$wave texture that is realized in 2D due to the absence of spin-orbit coupling. For the 6 H configuration in e), the bands show more splitting as we go away from the Fermi level. 
}
\label{Fig3}
\end{figure*}

In order to form an altermagnetic state, it is necessary to remove the $[C_{2,z}\parallel\mathcal{T}]$ symmetry while still maintaining a non-trivial spin group. The complete discussion can be found in the Discussion section, the main conclusion being that there are exactly two spin groups compatible with hydrogenated graphene. These are $^{2}m$, with a mirror plane perpendicular to the lattice (vertical) and paired with spin inversion, and $^{1}3^{2}m$, with a threefold rotational symmetry not paired with spin inversion and three vertical planes paired with spin inversion. As shown in Figure \ref{Fig3}a, we used STM manipulation to create a 4 H arrangement exhibiting the $^{2}m$ spin group, which can be explicitly verified by applying the corresponding transformations to the STM topography. As observed, applying a mirror symmetry to the STM image after time-reversal restores the system to its initial configuration. In contrast to the antiferromagnetic arrangement, the system lacks $[C_{2,z}\parallel\mathcal{T}]$ symmetry (see Supplementary Fig. 2). 
To validate our conjecture and confirm that this system exhibits the expected properties for an altermagnet, we conducted mean-field Hubbard tight-binding and DFT calculations, respectively included in 
Figure \ref{Fig3}b-d and Supplementary Figure 3. As predicted by Lieb’s theorem, the calculations show zero net magnetization for the system. Importantly, analysis in reciprocal space reveals spin-polarized energy bands along specific directions, as expected for an altermagnet. This behavior is consistently observed across various configurations of H atoms with equivalent symmetry properties, see Supplementary Figures 3 (DFT) and 4 (tight-binding). 
As revealed in Figure \ref{Fig3}d, the spin texture of the bands along the Brillouin zone exhibits a d-wave symmetry, with the centrosymmetry of the texture resulting from the absence of SOC. 
The remaining altermagnetic phase, with $^{1}3^{2}m$ spin group as predicted by the symmetry analysis in the Discussion section, requiring a configuration with at least 6 H atoms, is here demonstrated by the mean-field Hubbard tight-binding calculation, see Fig. \ref{Fig3}e-g. Notably, due to the reduced dimensionality, the spin texture exhibits an $i-$wave symmetry (P-6, in the notation of Ref. \cite{Jungwirth2022}, despite the fact that the spin group corresponds to a $g-$wave with B-4 in the analysis for 3D systems).  

\begin{figure*}
\centering 
\includegraphics[width=\textwidth]{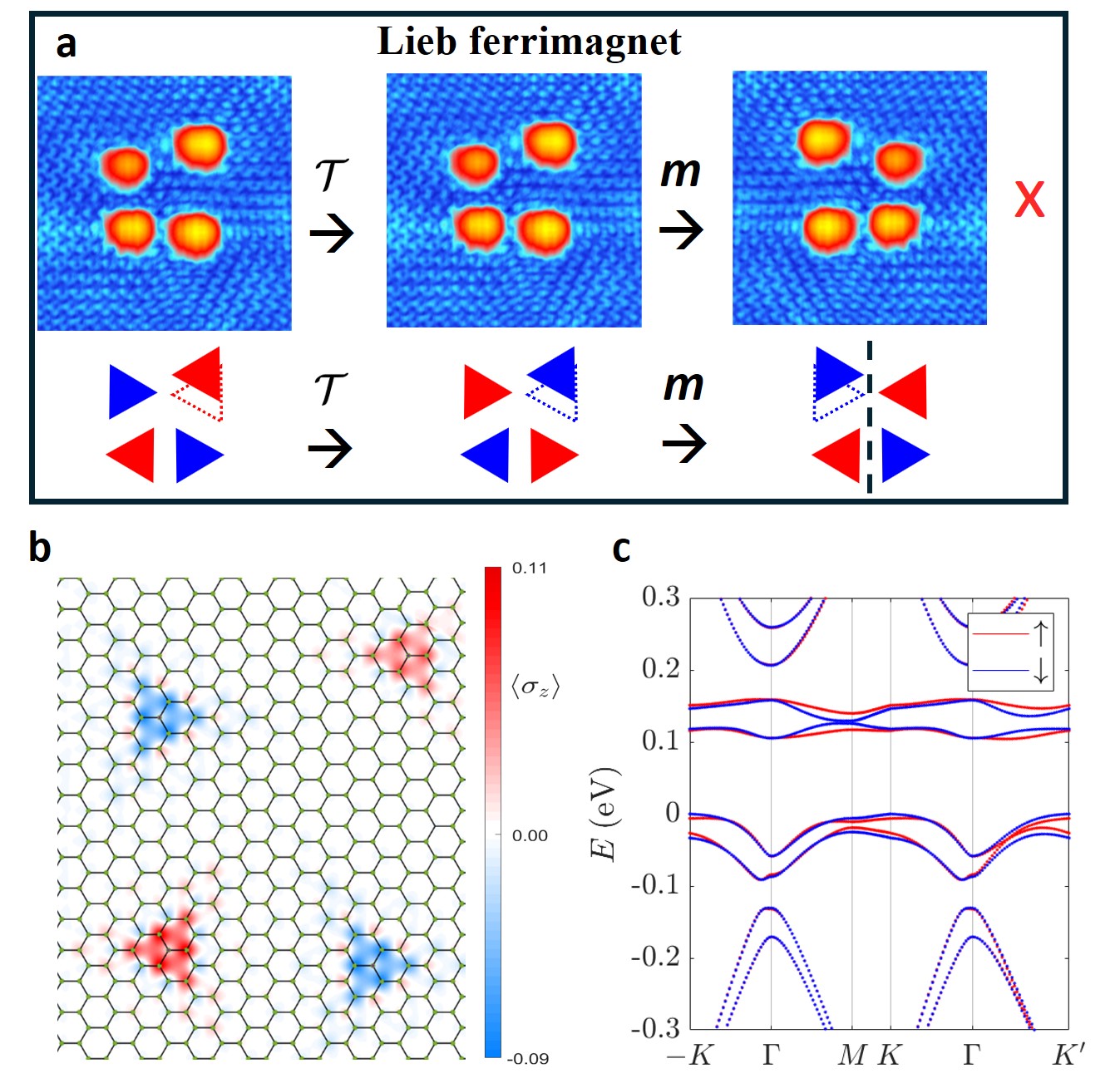}
\caption{\textbf{Lieb ferrimagnet.} a) Top, STM image (50mV, 0.1nA, 7x7nm$^2$) and symmetry transformations applied to the minimal experimental realization of a Lieb ferrimagnet. Bottom, triangle schematics and symmetry transformations. Dashed triangles highlight how crystal symmetry is broken. After $\mathcal{T}$ in spin space, there is no crystal operation that recovers the initial configuration. b) DFT magnetization of the experimental broken symmetry configuration, with zero net magnetization. c) Spin resolved electronic band structure of the configuration shown in b), The spin splitting is unrestricted due to the absence of any symmetry, and the perfect magnetization compensation is enforced by Lieb's theorem.  }
\label{Fig4}
\end{figure*}

In the absence of spatial transformations connecting to the initial configuration after a time-reversal operation, the system develops fully (or unrestricted) spin-polarized energy bands. 
As shown in Figure \ref{Fig1}, with less than 4 atoms achieving this full spin polarization of the energy bands is only possible in systems with a non-vanishing net magnetization ($S=1/2$ for 1 H atom, $S=1$ for 2 ferromagnetically arranged H atoms, and $S=1/2$ for 3 ferrimagnetically arranged H atoms). However, with 4 H atoms it becomes possible to combine zero net magnetization with spin polarized bands without any spatial symmetry connecting back to the initial state. We refer to these configurations as "Lieb ferrimagnets", drawing an analogy to existing fully compensated ferrimagnets~\cite{Zhang2015}. The key distinction lies in the fact that, in this case, the compensation of both spin states—resulting in a vanishing total magnetization—is
enforced by Lieb’s theorem rather than by a specific magnetic space group lacking space-time inversion symmetry. Additionally, the spatial symmetry in these configurations is broken by individual hydrogen atoms displaced from their symmetry positions in real space, as expected from random H arrangements, rather than requiring fine-tuned compositions or external conditions as is typical in traditional ferrimagnets~\cite{finley2020spintronics,Cai2020, Siddiqui2018, Caretta2018, Kawamura2024, Nayak2015,xu2023evolution}. Fig. \ref{Fig4} illustrates an experimental realization of the Lieb ferrimagnetic phase, achieved with 4 H atoms slightly displaced from an altermagnetic configuration. In this case, after a spin-flip, the system does not revert to its original configuration regardless of the subsequent spatial symmetry transformation applied to the experimental STM topography, see Fig. \ref{Fig4}a. Our DFT and mean-field Hubbard calculations confirm that the system exhibits zero net magnetization while maintaining fully spin-polarized energy bands, see Fig. \ref{Fig4}b-c and Supplementary Figure 5, respectively. This behavior has been consistently verified across numerous broken-symmetry, fully compensated configurations (see Supplementary Figure 6). 

Experimentally accessing the spin-splitting in these artificial structures is not straightforward. However, as opposed to antiferromagnets and altermagets, the full (asymmetric) splitting of the energy bands of compensated ferrimagnets should be reflected in the local density of states (LDOS) as a series of peaks slightly shifted in energy due to spin-splitting, as shown by our DFT and tight-binding calculations (see Supplementary Fig. 7). Correspondingly, our experimental \textit{dI/dV}  measurements on individual hydrogen atoms in the Lieb-ferrimagnetic configurations reveal multiple peaks slightly shifted in energy, as expected for spin-split states (Supplementary Fig. 7i-l). It is however challenging to definitively attribute these shifts to actual spin-splitting because of the coexistence of elastic and inelastic contributions to the tunneling current, associated to spin-split states and spin-flip excitation processes respectively.

Importantly, as shown in Supplementary Figure 5d, our mean-field tight-binding calculations demonstrate that the spin polarization of the bands in this new magnetic phase is robust against small levels of doping. This enables the long-sought ability to tune the spin polarization of currents through external electrostatic doping. In Supplementary Figure 8, we summarize the three possible magnetically-compensated structures that can be created in graphene with 4 H atoms.

\section{Discussion}
The spin degeneracy of the band structure across the whole Brillouin zone, sometimes referred to as Kramer's degeneracy, is enforced by space-time inversion symmetry $[\mathcal{P}\parallel\mathcal{T}]$, the operation to the left (right) of the double bar acts exclusively in real (spin) space, $\mathcal{P}$ represents spatial inversion and $\mathcal{T}$ represents time-reversal or, in the context of spin groups, a general inversion of spin. In non-relativistic collinear magnets, the spin-flip associated to $\mathcal{T}$ can always be realized by a $C_{2}$ pure-spin rotation along an axis perpendicular to all magnetic moments. In 2D materials, Kramer's degeneracy is analogously imposed by $[C_{2,z}\parallel\mathcal{T}]$ symmetry. Upon adsorption of an arbitrary number of H atoms
on the graphene lattice, $\mathcal{P}$ symmetry is always removed (assuming that all the adsorption occurs on one side of the lattice), and $C_{2,z}$ symmetry is also removed in general except in some particular cases. Specifically, $[C_{2,z}\parallel\mathcal{T}]$ is preserved for 2 H atoms if and only if they are adsorbed on different sublattices, and for 4 H atoms if and only if they form a parallelogram (with the rotation axis containing its centroid), with 2 H in each sublattice. The electronic structure of these specific configurations is Kramer's degenerate, the remaining systems exhibiting a certain spin-splitting which is expected to be sizable only in the presence of non-vanishing magnetic orderings, since the strength of SOC is remarkably small in our system of light elements. Here, the local magnetic ordering arises due to the unpaired electron mechanism, as previously described. Concomitantly, by Lieb's theorem \cite{Lieb1989} any configuration with an evenly distributed number of H atoms among both sublattices will present a vanishing total magnetic moment in the ground state. It can therefore be concluded that a graphene lattice with an even number of adsorbed H atoms equal to or greater than 4, distributed evenly among both sublattices, will in general exhibit a ground state with perfectly compensated, collinear magnetism and spin-split bands. The only exceptions are H arrangements with a center of inversion, namely those for which the H atoms form a zonogon.

\subsection{Spin groups in hydrogenated graphene altermagnets}
While the default phase for an evenly distributed arrangement of H atoms will most probably correspond to a Lieb ferrimagnet, it is however possible to engineer altermagnetic states by placing the H atoms in symmetry-related sites, in reminiscence of the antiferromagnetic construction with 4 H shown in Figures \ref{Fig2}b-d. Altermagnets are essentially classified according to their spin point group. There are 37 possibilities in 3D systems (not all of them compatible with lower dimensionalities, and no more options appear in 2D), with the 10 ones that contain inversion named Laue groups. The incoming discussion can be followed with the aid of, for example, Table S.I of Ref. \cite{Jungwirth2022} (see Supplementary Information therein), where we note that a superscript $2$ ($1$) preceding an operation in the Hermann–Mauguin notation of point groups denotes the pairing (or not) with spin inversion. First, we exclude all cubic and tetragonal groups, which are respectively not compatible with 2D and graphene's hexagonal lattice. Furthermore, we disregard all groups which include the inversion of $z$, due to the presence of H on a single side of the lattice, which is not an essential limitation in any case (if one thinks about lifting such constraint by considering vacancies instead of H atoms) since these operations do not affect reciprocal space and there always exists a partner group that excludes such operations. 
From the observation that there is a one-to-one correspondence between sublattice of adsorption and spin orientation, it follows that any operation that permutes (does not permute) the sublattices would necessarily have to be paired (not be paired) with the spin inversion operation in order to be a symmetry of the hydrogenated graphene system. The remaining operations that permute the sublattices are $C_{2,z}$, $C_{6,z}$ and (vertical) mirror planes bisecting the carbon bonds, while the operations that do not permute the sublattices are $C_{3,z}$ and the (vertical) mirror planes passing through carbon atoms. However, from these possibilities it is still necessary to discard the groups containing $C_{2,z}$ (and thus also $C_{6,z}$) in order to avoid Kramer's degeneracy. Combining all these arguments, we conclude that the only possible spin groups that are compatible with our platform are $^{2}m$, with the mirror plane perpendicular to the lattice, and $^{1}3^{2}m$, in its only possible orientation (threefold axis perpendicular to the lattice). Both are non-Laue spin groups, respectively presenting $d$ and $g-$wave parities or textures in 3D reciprocal space \cite{Jungwirth2022}, however, the latter is further subdivided within the planar Brillouin zone by the non-relativistic centrosymmetry and actually results in an $i-$wave-like symmetry. It is clear that the former group can be realized with any even number of H atoms excluding 2 (which always preserves $[C_{2,z}\parallel\mathcal{T}]$, assuming opposite sublattices), while the latter requires a multiple of 6 (which is difficult to realize in practice for 12 or more). Both of these distinct altermagnetic configurations are shown in Figure \ref{Fig3}.


In conclusion, we have introduced a bottom-up strategy to engineer magnetic structures that simultaneously exhibit perfect spin compensation and lifted spin degeneracy, using graphene and individual hydrogen atoms as building blocks. Owing to Lieb’s theorem, this approach can be scaled up to a large number of hydrogen atoms while maintaining zero net magnetization, provided both graphene sublattices are equally populated. We have shown that configurations with (spatial) vertical mirror symmetry serve as fundamental units of altermagnets, whereas those lacking such symmetry yield Lieb-ferrimagnets. In both scenarios, $[C_{2,z}\parallel\mathcal{T}]$ symmetry is broken, leading to a spin-polarized band structure coexisting with zero net magnetization. By enabling direct visualization and manipulation of spatial and spin symmetries at the atomic scale, our technique opens new avenues for the targeted design of magnetic phases. This concept can be extended to other substrates or applied through self-assembly techniques to realize altermagnets from molecular arrays on surfaces. Additionally, the required symmetry conditions may be tailored via strain or twisting in two-dimensional materials, facilitating the creation of both altermagnets and fully spin-split, compensated ferrimagnets. 
 
\section{Methods}

\subsection{Sample preparation and STM measurements}

Multilayer graphene is grown epitaxially in ultra-high vacuum (UHV) conditions by thermal decomposition of a 6H-SiC(000-1) sample \cite{PhysRevB.77.165415}. In this system, due to the rotational disorder the upmost graphene layer is neutral and electronically decoupled, behaving essentially as free-standing graphene \cite{PhysRevLett.100.125504}.

We deposit atomic H on graphene by thermal dissociation of a beam of H$_2$ using a hot W filament~\cite{Gonzalez-Herrero2016, Hornekar2006}. H$_2$ pressure is regulated by a leak valve and fixed to $4 \cdot 10^{-7}$ Torr as measured in the preparation chamber. The filament is held at 1900K for 6 minutes with the sample placed 10 cm away at room temperature. During the whole process---imaging pristine graphene sample => depositing H atoms on it => and imaging it back---the sample was maintained in the same UHV system.

All experiments were performed on a homemade STM at a temperature of 4K. \textit{dI/dV} conductance curves were obtained by numerical differentiation of measured \textit{I-V} curves. The STM data was acquired and processed using the WSxM software~\cite{Horcas2007}.\\

\subsection{Atomic H manipulation}  

Different H-atom arrangements are created by atomic manipulation using the STM tip. Selected graphene regions are “cleaned” by removing, and thus collecting, all H atoms with the tip. This collection can be carried out either atom-by-atom---by contacting each H atom individually at low voltages---or collectively, by rapidly scanning the surface at low bias voltages and high currents, (5-100mV and 5-10nA). The collected H atoms are deposited on a selected graphene region by applying negative sample voltages pulses (up to -9V).  A slightly variable number of H atoms is thus deposited with nm scale precision. To create the intended design, with the desired total magnetization and symmetry, excess H atoms are removed one by one, approaching the STM tip (1-2 \AA) to their adsorption site, leaving only the desired pattern. At present, the deposition step is subject to some incertitude, determined by the specific tip apex termination. However, the selective removal of single H atoms works with essentially a 100\% efficiency, enabling the precise construction of the selected symmetries.

\subsection{Density functional theory results}
The self-consistent electronic Hamiltonian was obtained employing the massively parallel version of the \texttt{CRYSTAL} code \cite{crystal23}, in all cases with the standard PBE exchange-correlation functional \cite{PBE}. We employed Gaussian basis sets consisting of a reduced pob-DZVP-rev2 \cite{vilela2019bsse} (removing the $d-$type functions) for carbon, suitable for the description of a bare graphene lattice, and an unmodified pob-TZVP \cite{peintinger2013consistent} for hydrogen. A single geometry optimization was performed with one H atom in a $12\times12$ supercell, restricted to the carbon atom in the adsorption site, its 3 nearest neighbors and the H atom; resulting in an attraction of the carbon cluster towards the H atom and in particular a C-H bond length of $1.13$\AA, with the central carbon lifted $0.38$\AA$ $ above the lattice. These displacements were individually applied for each H region in subsequent calculations, which were carried out in a $24\times24$ supercell to minimize spurious interactions within the computational capabilities. Experimental configurations are rescaled to fit the supercell size, keeping the proportions of the shape defined by the H vertices as unaltered as possible. A threshold of $10^{-10}$ for overlap and penetration of Hartree integrals (see Ref. \cite{crystal23Manual}) was employed, in conjunction with an uniform sampling of 20 points in the irreducible Brillouin zone of the supercell. An initial magnetic guess was applied to the first 3 carbon neighbors of each adsorption site, and the total initial spin was kept constant during 35 self-consistency cycles (always below $20\%$ of the final number of iterations). A tolerance of $10^{-7}$ Hartree was set to define the end of the self-consistency, after observing no significant changes from $10^{-9}$ Hartree in test systems. 
The resulting atomic magnetic moments are interpolated with a biharmonic spline method \cite{sandwell1987biharmonic} as implemented in \texttt{MATLAB}. Band structures and atom-projected densities of states are computed manually from the Hamiltonian and overlap matrices, of dimension $10392$ for each spin block.

\subsection{Mean-field Hubbard model}

The conduction band, consisting of $p_z$ electrons, of graphene sheet chemisorbed with hydrogen atoms can be described by the following mean-field Hubbard Hamiltonian~\cite{JJPalacios2008}
\begin{align}
	\mathcal{H} = & t \sum_{<i, j>} \sum_\sigma c_{i,\sigma}^\dagger c_{j,\sigma} + U \sum_i \left[ n_{i,\uparrow} \langle n_{i,\downarrow} \rangle + n_{i,\downarrow} \langle n_{i,\uparrow}\rangle\right]\nonumber\\
	& + \epsilon_H \sum_{i\in\text{\{C-H sites\}}} \sum_\sigma n_{i,\sigma} 
\end{align}
where $c_{i,\sigma}^\dagger$ ($c_{i,\sigma}$) is the creation (annihilation) operator of an electron with spin $\sigma$ at $i^{\text{th}}$ site of the graphene lattice. The first term of the Hamiltonian is the nearest-neighbour hopping term, where parameter $t$ is the hopping parameter and $<i,j>$ signifies pair of nearest-neighbour sites. The second term is the mean-field Hubbard term describing the on-site Coulomb repulsion between two electrons (which will have opposite spins) present on the same site, where U is the Hubbard potential, $n_{i\sigma} = c_{i\sigma}^\dagger c_{i\sigma}$ is the number operator, and $\langle n_{i,\sigma} \rangle$ is the ground state expectation value of the number operator calculated self-consistently. Since the sites on which hydrogen atoms are chemisorbed (C-H sites) are seen as vacancies in the lattice by the $p_\pi$ electrons, we introduce an on-site potential $\epsilon_H = U/2+100t \gg 6t$ (the bandwidth for pristine graphene) through the third term to simulate vacancies. The calculation presented in Figure \ref{Fig3} was done on a $30\times 30$ supercell with $U=2t$ and $t=2.5$ eV.

\section{Data availability} 

The data supporting the findings of this study are available from the corresponding authors upon reasonable request. 

\bibliography{biblio} 

\begin{acknowledgement}
We acknowledge helpful discussions with J. Fernández-Rossier. Graphene samples are provided by P. Mallet and J-Y. Veuillen.

We acknowledge financial support from the Spanish Ministry of Science and Innovation, through projects  (grants nos. PID2023-149106NB-I00, TED2021-131323B-I00, and PID2022-141712NB-C21), the Mar\'ia de Maeztu Program for Units of Excellence in R\&D (grant no. CEX2023–001316-M), the Comunidad de Madrid and the Spanish State through the Recovery, Transformation and Resilience Plan [Materiales Disruptivos Bidimensionales (2D), (MAD2DCM)-UAM Materiales Avanzados] and the NMAT2D-CM program under grant S2018/NMT-4511, the European Union through the Next Generation EU funds, the Generalitat Valenciana through the Program Prometeo (2021/017). M. A. García-Blázquez acknowledges financial support from Universidad Autónoma de Madrid through a FPI-UAM grant. S. Chourasia acknowledges financial support from grant PREP2022-000250 funded by MICIU/AEI/10.13039/501100011033 and by ESF+. B. Viña-Bausá acknowledges funding from the Spanish Ministerio de Universidades through the PhD scholarship No. FPU22/03675. The authors thankfully acknowledge Red Española de Supercomputación for the computational resources provided by Universidad de Málaga through the projects FI-2024-1-0038, FI-2024-2-0016 and FI-2024-3-0010.


\end{acknowledgement}

\section{Author contributions}
B.V. carried out the main body of measurements and the elaboration of the figures, M.A.G.B. carried out the DFT calculations and the group theory analysis, S.C. carried out the mean-field Hubbard calculations, R.C. and D.E. contributed to the experimental measurements, I.B. coordinated the project, planned the structure of the manuscript and wrote the main body of the paper, J.J.P. had the original idea to construct altermagnets and also coordinated the project. B.V., M.A.G.B. and J.J.P. strongly contributed to the writing with the input of all authors.

\section{Competing interests}
The authors declare no competing financial interests.

\section{Figures}

\renewcommand{\figurename}{Extended Figure}
\setcounter{figure}{0}
\begin{figure*}
\centering 
\includegraphics[width=\textwidth]{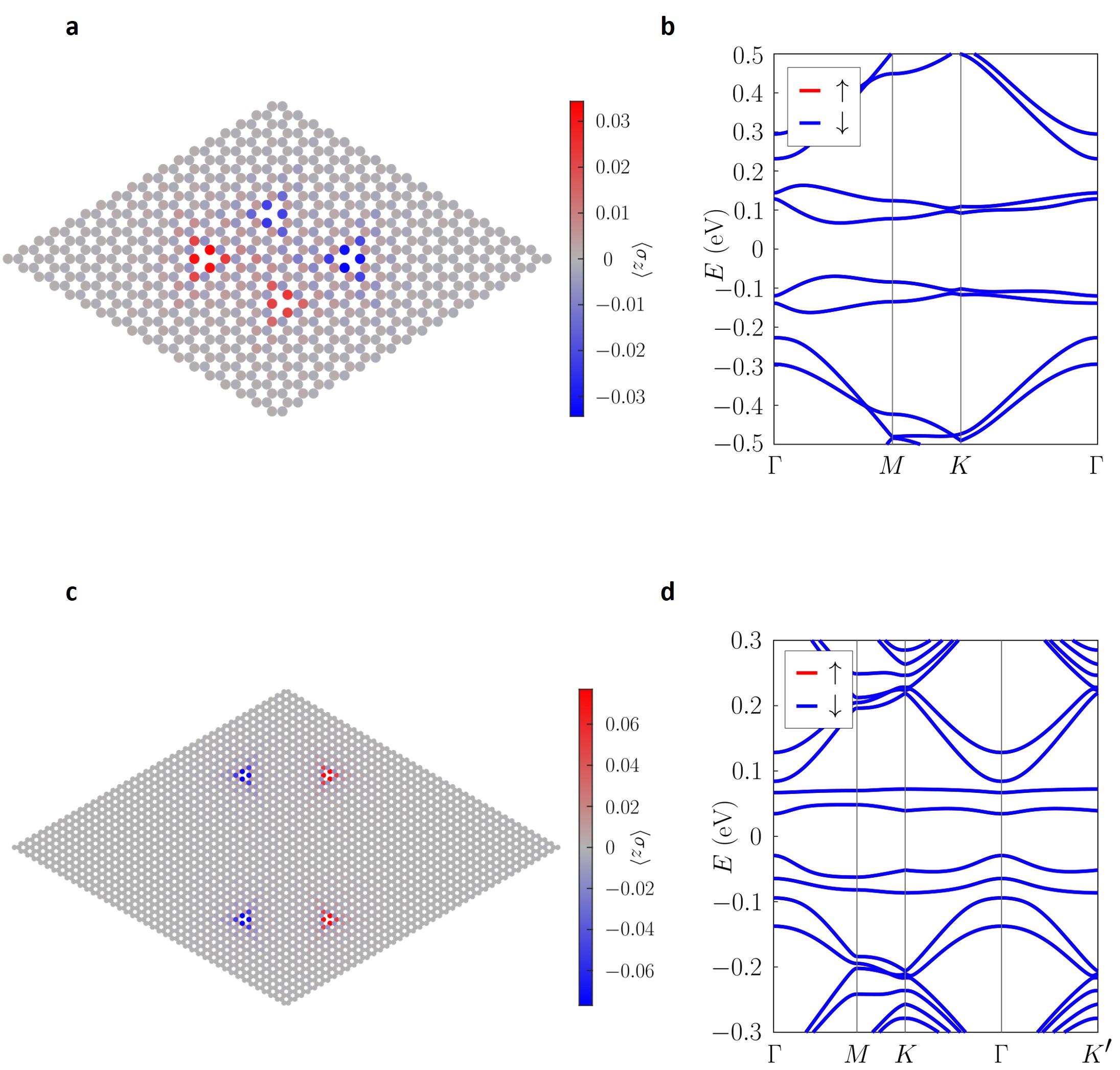}
\caption{\textbf{Tight-binding calculations of two different 4 H antiferromagnetic configurations} (i.e. with parallelogram symmetry). Magnetization (a,c) and corresponding spin resolved band structure showing Kramer’s degeneracy (b,d).}
\label{SFig1}
\end{figure*}

\begin{figure*}
\centering 
\includegraphics[width=\textwidth]{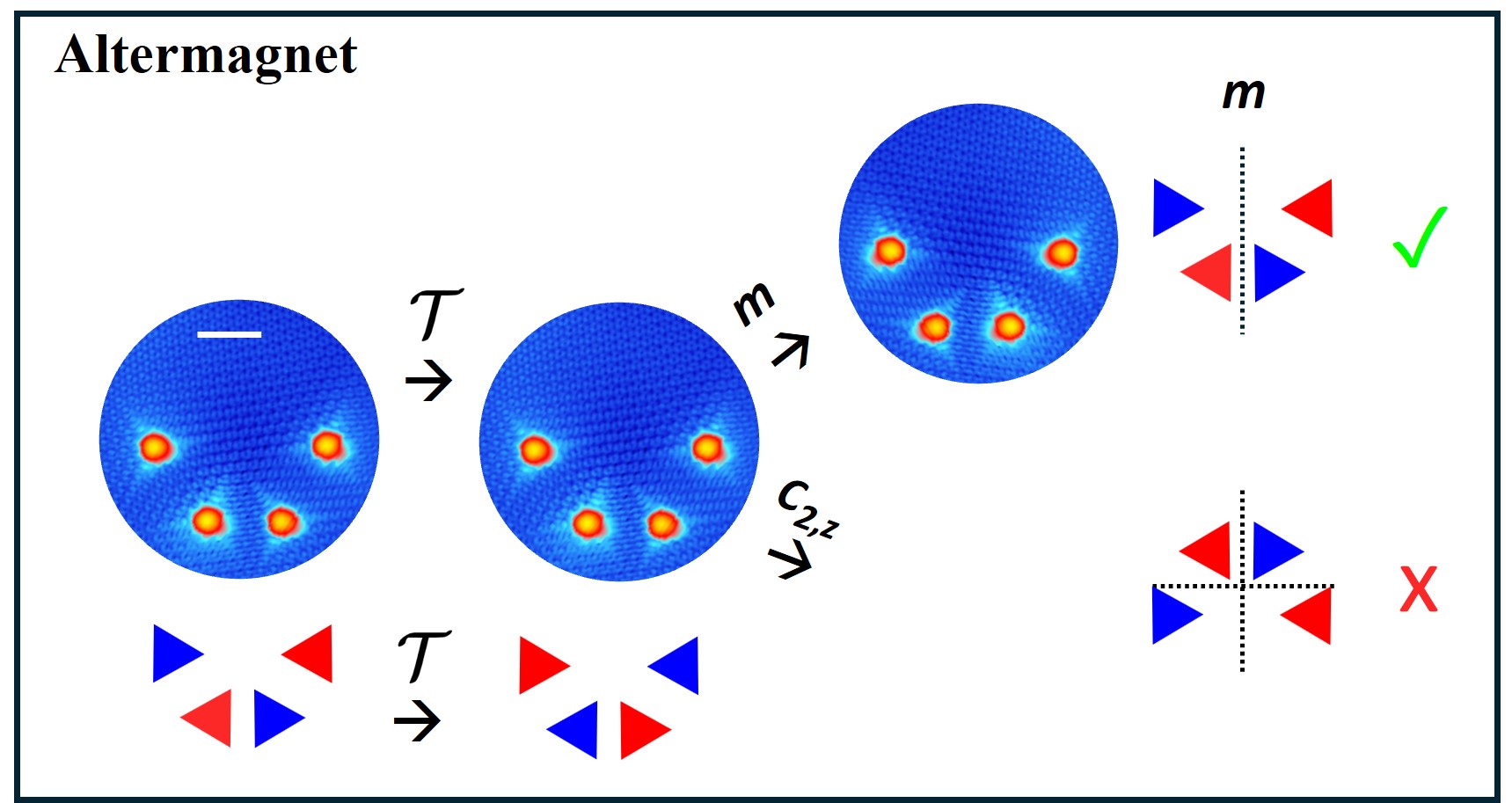}
\caption{\textbf{Altermagnetic configuration with 4 H atoms} (120mV, 0.1nA, scalebar=2nm). The symmetry requirements are verified on the STM image. After performing a time reversal on spin space ($\mathcal{T}$), a mirror plane (\textit{m}) recovers the original configuration, while $C_{2,z}$ is broken.}
\label{SFig2}
\end{figure*}

\begin{figure*}
\centering 
\includegraphics[width=\textwidth]{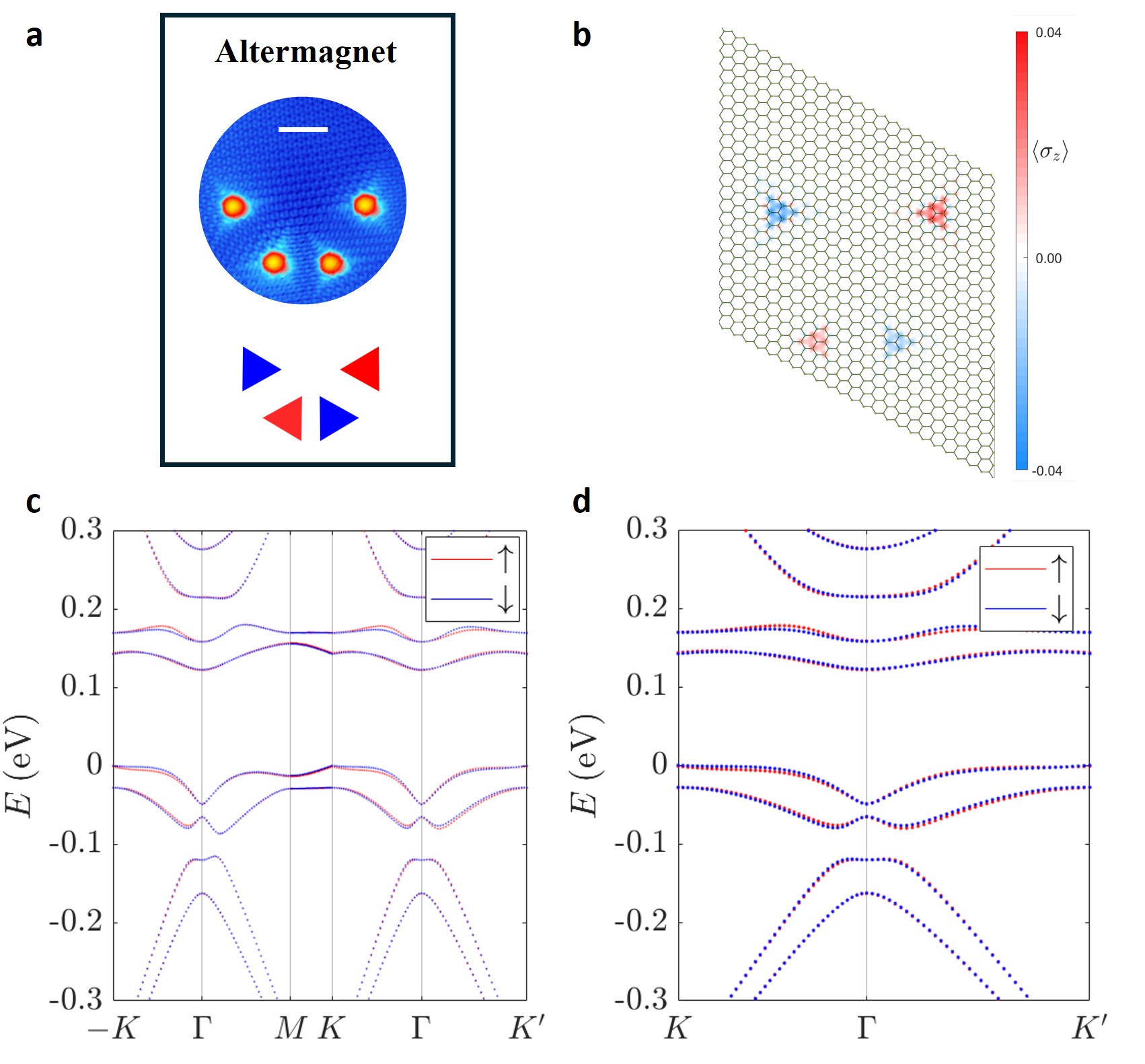}
\caption{\textbf{DFT calculation of 4 H graphene as an altermagnet.} a) Experimental configuration and symmetry schematics of a 4 H arrangement in graphene with spin group containing a single vertical mirror plane. b) DFT calculated magnetization of a similar altermagnetic configuration. c) Corresponding energy bands d) Zoom of the band structure. STM image: 120mV, 0.1nA, scalebar=2nm.
}
\label{SFig3}
\end{figure*}

\begin{figure*}
\centering 
\includegraphics[width=\textwidth]{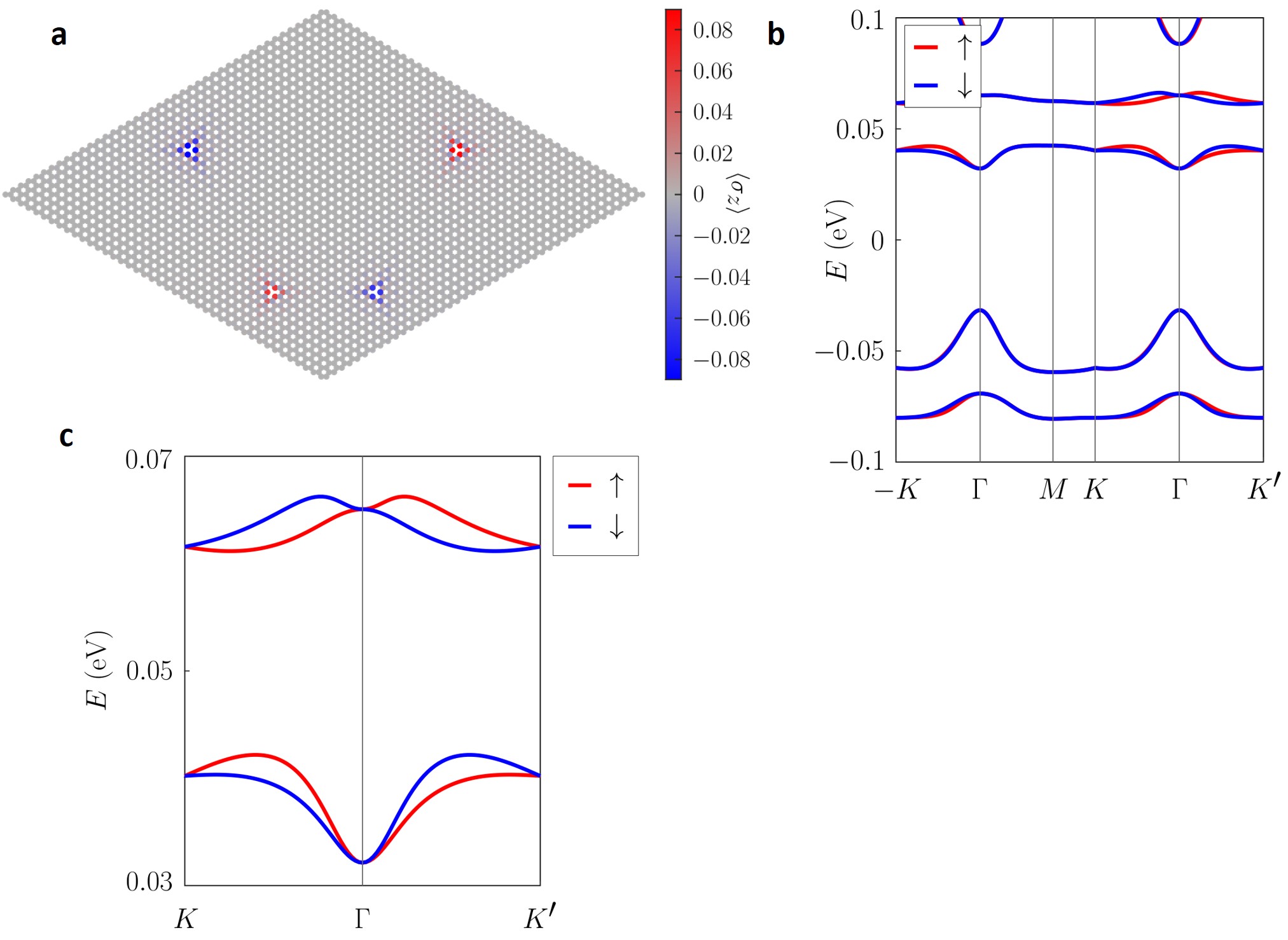}
\caption{\textbf{ Tight-binding calculation of a different 4 H altermagnet on graphene. } Magnetization (a) and energy bands (b and zoom in c) for a different 4 H arrangement in graphene with mirror symmetry and, again, altermagnetic properties.}
\label{SFig4}
\end{figure*}

\begin{figure*}
\centering 
\includegraphics[width=\textwidth]{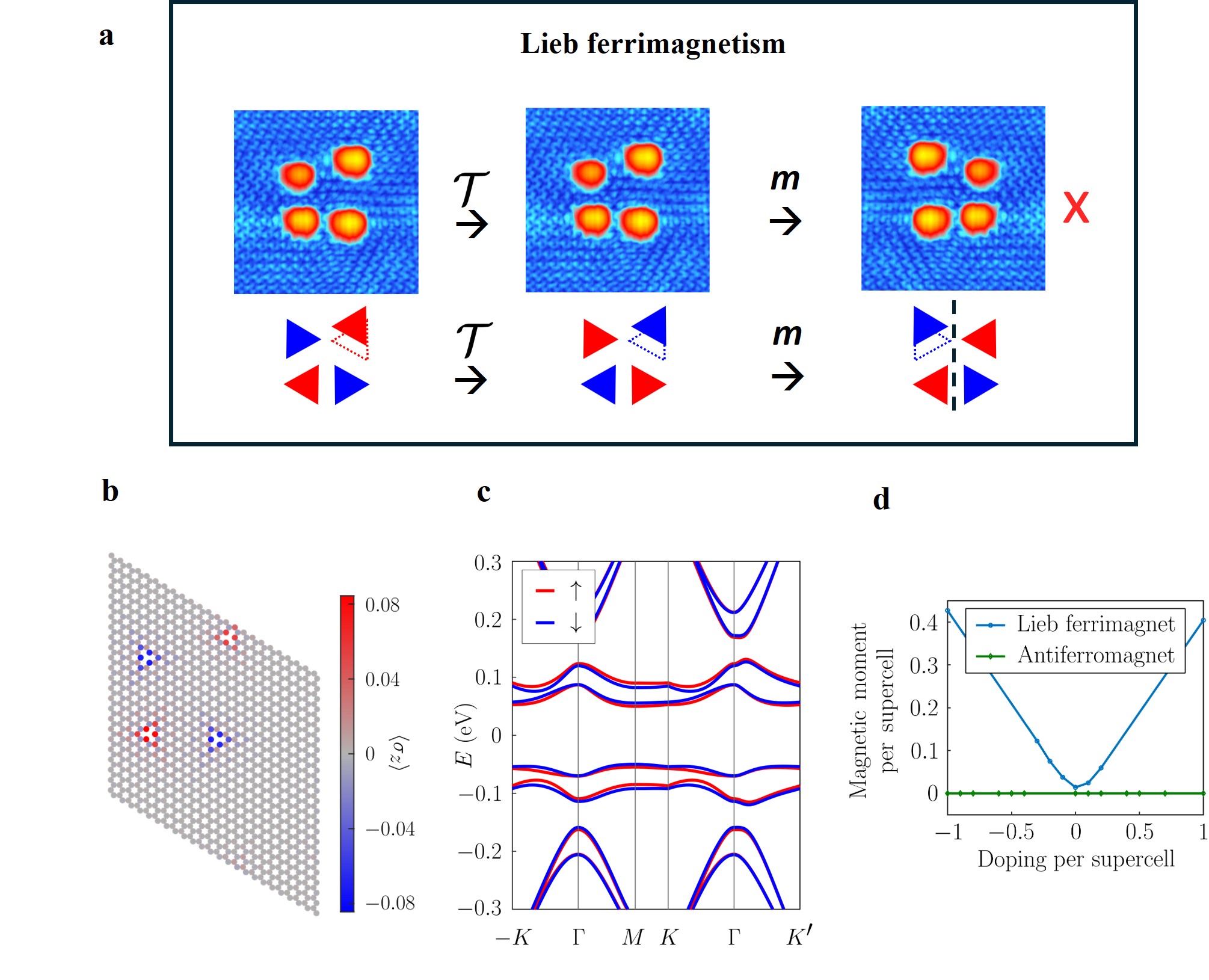}
\caption{\textbf{Tight binding calculations for a Lieb ferrimagnet.} a) Top, STM image and symmetry transformations applied to the minimal experimental realization of a Lieb ferrimagnet. Bottom, triangle schematics and symmetry transformations. Dashed triangles highlight how crystal symmetry is broken. After $\mathcal{T}$ in spin space, there is no crystal operation that recovers the initial configuration. b) Tight binding magnetization of the broken symmetry configuration [shown in a) top left] with zero net magnetization. c) Spin resolved electronic band structure of the configuration shown in b) having unrestricted spin-splitting due to the absence of any symmetry and perfect magnetization compensation enforced by Lieb's theorem. d) Dependence of net magnetic moment per supercell, $\sum_i [\langle n_{i\uparrow} \rangle - \langle n_{i\downarrow} \rangle]/2$, on doping (number of electrons added to the half-filled state of the supercell) for a Lieb ferrimagnet similar to b) but scaled down to a $12\times12$ supercell and a corresponding centrosymmetric antiferromagnet obtained by rearranging the positions of the H atoms. For the antiferromagnet, most of the solutions obtained after doping were non-magnetic due to small distances between the H atoms but a similar calculation on a bigger supercell is expected to be magnetic with net zero magnetic moment after doping.}
\label{SFig5}
\end{figure*}

\begin{figure*}
\centering 
\includegraphics[width=380pt]{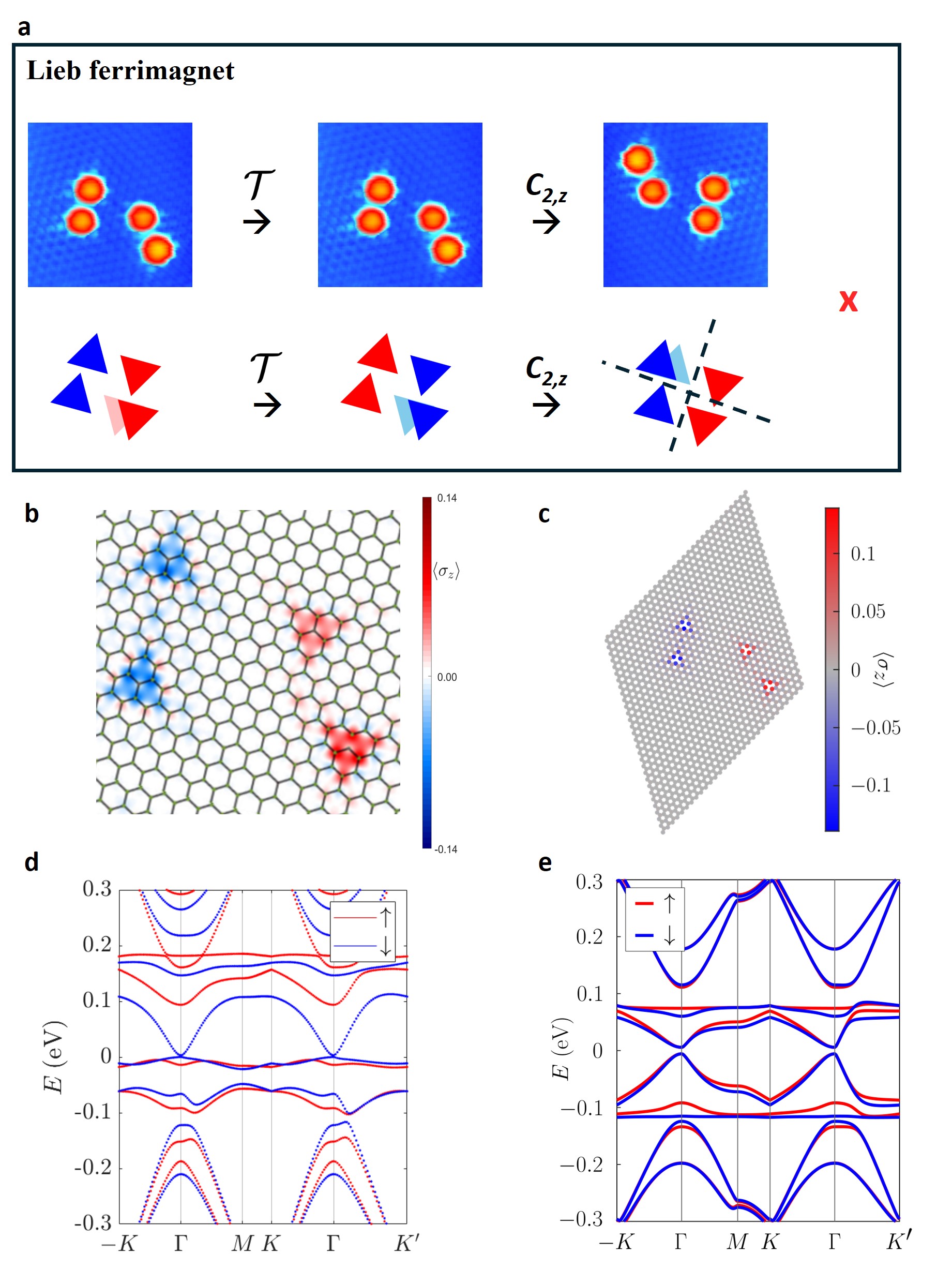}
\caption{\textbf{Experiment and theory of another 4 H Lieb ferrimagnet on graphene.} a) STM image (100mV, 0.1nA, 6.8x6.8nm2 ) and symmetry transformations applied to a different experimental realization of a Lieb ferrimagnet. b,c) Calculated magnetization by DFT and tight binding respectively. d,e) Spin resolved electronic bands calculated by DFT and tight binding respectively.}
\label{SFig6}
\end{figure*}

\begin{figure*}
\centering 
\includegraphics[width=\textwidth]{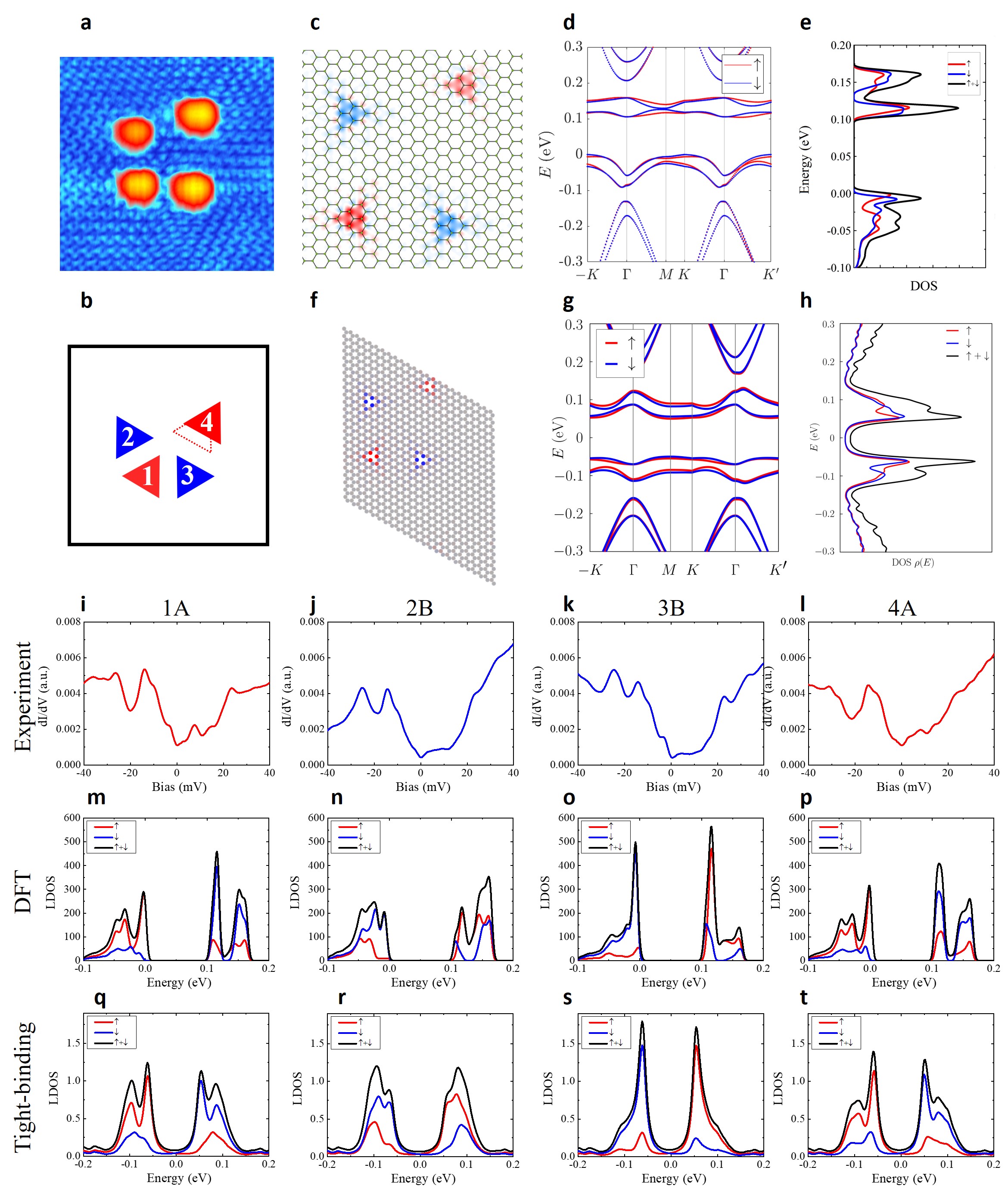}
\caption{\textbf{ Experiment and theory of 4 H Lieb Ferrimagnet on graphene.} a) STM image (50mV, 0.1nA, 7x7nm$^2$) of the experimental realization of a Lieb ferrimagnet. b)Triangle schematics of the configuration. DFT (c,d,e) and tight binding (f,g,h) magnetization, band structure and DOS respectively of the experimental configuration shown in a). (i-l) Experimental \textit{dI/dV} measured on each H atom of the configuration. Control parameters: 50mV, 0.1nA. DFT (m-p) and tight biding (q-t) LDOS projected on each H site. }
\label{SFig7}
\end{figure*}

\begin{figure*}
\centering 
\includegraphics[width=\textwidth]{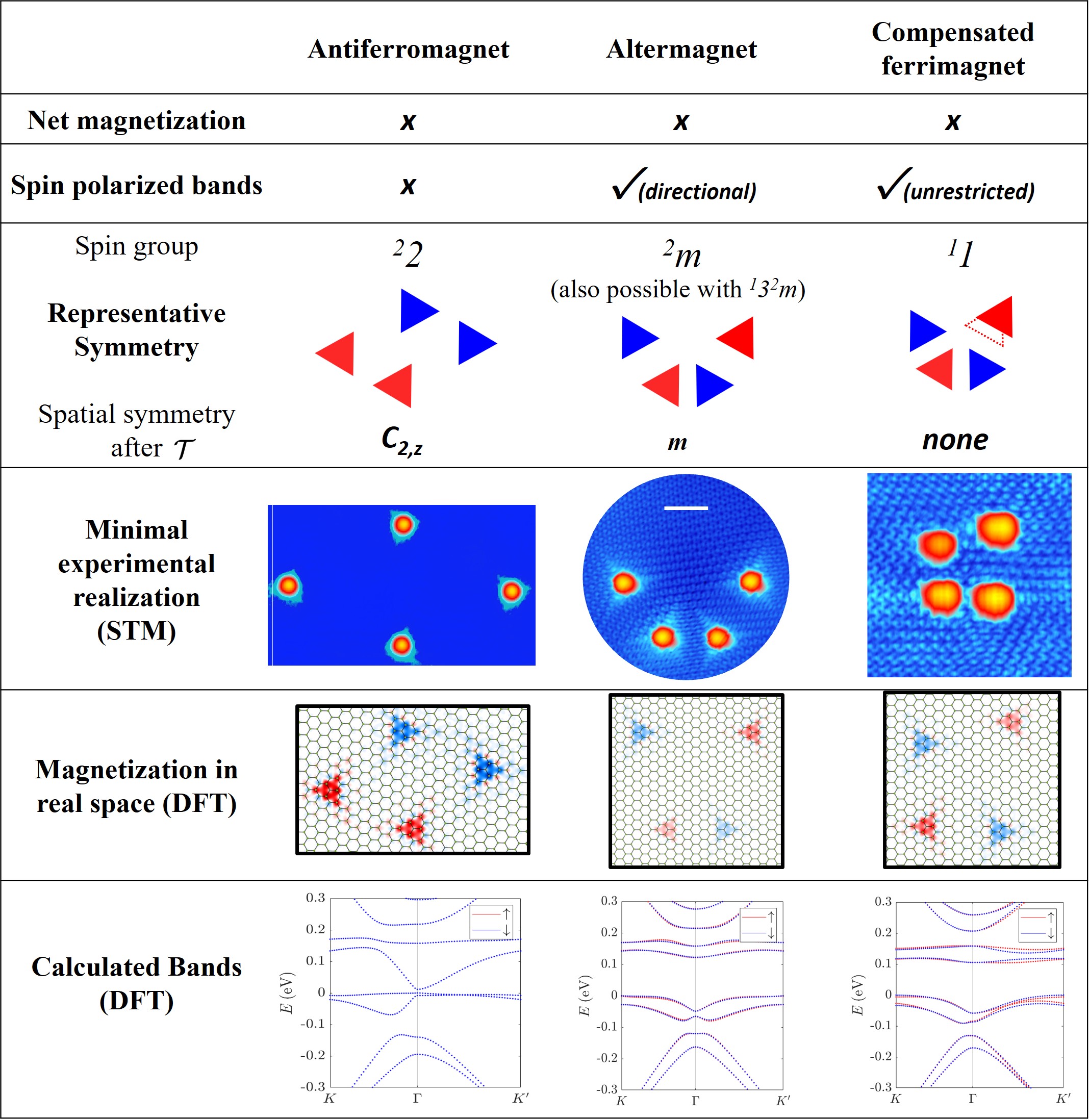}
\caption{\textbf{Building unconventional magnetic phases with 4 H atoms in graphene.}  Summary of all possible collinear, non-relativistic, magnetic phases that can be  built on graphene with 4 H atoms and zero net magnetization (even sublattice distribution). By selectively adjusting or breaking the symmetry of the configuration, it is possible to obtain spin polarized bands, in coexistence with a zero net magnetization enforced by Lieb's theorem. The fourth and fifth rows show the STM image and DFT calculated magnetization of each magnetic phase. The last row shows the spin resolved electronic bands. Configurations that fulfill $\mathcal{T}$ + $C_{2,z}$ symmetry correspond to an antiferromagnet, rendering Kramer's degenerate bands; those with $\mathcal{T}$ + \textit{m} to a d-wave altermagnet, with directional splitting of the bands; and $\mathcal{T}$ broken configurations correspond to fully compensated ferrimagnets, with fully spin-split bands. An alternative i-wave altermagnetic state can also be realized employing 6 H atoms. STM image parameters by columns: (80mV, 0.1nA, 16x25nm$^2$); (120mV, 0.1nA, scalebar=2nm) ; (30mV, 0.1nA, 7x7nm$^2$).}
\label{SFig8}
\end{figure*}

\end{document}